\documentclass[twocolumn,showpacs,preprintnumbers,amsmath,amssymb]{revtex4}

\usepackage{graphicx}
\usepackage{dcolumn}
\usepackage{bm}

\usepackage{epsfig}
\usepackage{amsmath}
\usepackage{amssymb}
\usepackage{amsfonts}
\usepackage{color}

\allowdisplaybreaks     

\newcommand{\old}[1]{{\textcolor{green}{ }}}

\renewcommand{\vec}[1]{\mathbf{#1}}

\begin{document}

\preprint{APS/123-QED}

\title{Time-dependent Internal DFT formalism and Kohn-Sham scheme.}

\author{J{\'e}r{\'e}mie Messud}
\affiliation{
Universit\'e de Toulouse; UPS; Laboratoire de Physique
  Th\'eorique (IRSAMC); F-31062 Toulouse, France
}
\affiliation{
CNRS; LPT (IRSAMC); F-31062 Toulouse, France
}

\date{\today}

\begin{abstract}
We generalize to the time-dependent case the stationary Internal DFT / Kohn-Sham formalism presented in Ref. \cite{Mes09}.
We prove that, in the time-dependent case, the internal properties of a self-bound system (as an atomic nuclei or a Helium droplet)
are all defined by the internal one-body density and the initial state.
We set-up a time-dependent Internal Kohn-Sham scheme as a practical way to compute the internal density. The main difference with the traditional DFT / Kohn-Sham formalism
is the inclusion of the center-of-mass correlations in the functional.
\end{abstract}


\maketitle

\section{Introduction.}

Traditional Density Functional Theory (DFT) \cite{Dre90,Koh99,DFTLN} and its time-dependent generalization \cite{Run84,Gro94}
have evolved into standard tools for the description of electronic properties in condensed-matter physics and quantum chemistry
through the simple local density instead of the less tractable $N$-body wave function.
Stationary DFT is based on the Hohenberg-Kohn (HK) theorem \cite{Hoh64}, which
proves that, for any non-degenerate system of N Fermions
or Bosons \cite{Dre90} put into a local external potential, the $N$-body ground-state wave function can be written as a functional of the local ground-state density.
A similar theorem exists for the time-dependent case \cite{Run84,Gro94}, where a dependence on the initial state appears.
The Kohn-Sham (KS) scheme \cite{Koh65} and its time-dependent generalization \cite{Run84,Gro94} provide a straightforward method 
to compute self-consistently the density in a quantum framework,
defining the non-interacting system (i.e.\ the local single-particle potential) which reproduces the exact density.

Traditional DFT is particularly well suited to study the electronic properties in molecules \cite{Kre01}.
As a molecule is a self-bound system, the corresponding Hamiltonian is translationally invariant 
(which ensures Galilean invariance of the wave function \cite{foot1}),
and one can apply the Jacobi coordinates method.
This permits to decouple the center-of-mass (c.m.) properties from the internal ones,
and to treat correctly the redundant coordinate problem
(i.e.\ the fact that one coordinate is redundant for the description of the internal properties \cite{Schm01a})
and the c.m. correlations.
But as the nuclei are much heavier than the electrons,
we can apply the Jacobi coordinates method to the nuclei only,
so that only the nuclei will carry the c.m. correlations,
and use the clamped nuclei approximation.
Then, one recovers the ``external'' potential of traditional DFT, of the form $\sum_{i=1}^N v_{ext}(\mathbf{r}_i)$,
which accounts for the nuclear background as seen by the electrons in the frame attached to the c.m. of the nuclei.
Thus, traditional DFT is particularly adapted to the study of the electronic properties in molecules \cite{Kre01}.
It is implicitely formulated in the nuclear c.m. frame \cite{foot2} and the energy functional 
does not contain any c.m. correlations.
Of course, contrary to the whole molecule, the pure electronic system is not a self-bound system:
the $v_{ext}$ potential breaks translational invariance and is compulsory in order to reach bound states in the stationary case.

For other self-bound systems, as isolated atomic nuclei or He droplets,
the situation is intrinsically different because the masses of all the particles (Fermions or Bosons) are of the same order of magnitude.
As a consequence, to decouple the c.m. properties from the internal ones,
one has to apply the Jacobi coordinates method
to \textit{all} the particules.
The redundant coordinate problem (thus the c.m. correlations) will now concern all the particles and should be treated properly.
If a DFT exists, the c.m. correlations 
should be taken into account in the functional.

Moreover, no "external" potential of the form $\sum_{i=1}^N v_{ext}(\mathbf{r}_i)$ can be justified in the corresponding self-bound Hamiltonians
(we denote $\mathbf{r}_i$ the $N$ 
particules
coordinates related to any inertial frame as the laboratory).
One may be tempted to formulate a DFT using the traditional DFT conclusions in the limit $v_{ext} \to 0$,
but this would lead to false and incoherent results because:
\begin{itemize}
\item in the stationary case, the Hohenberg-Kohn theorem is valid only for external potentials that lead to bound many-body states \cite{Lie83},
which is not the case anymore at the limit $v_{ext} \to 0$ 
for translational invariant particle-particle interactions \cite{Mes09};
\item the form of $v_{ext}$ is not translationally invariant, but translational invariance is a key feature of self-bound systems \cite{Schm01a,Pei62,Mes09};
\item traditional DFT concepts as formulated so far are not applicable in terms of a well-defined internal density $\rho_{int}$, i.e.\ the density relative 
to the system's c.m.\ , which is of experimental interest \cite{Kre01,Eng07,Mes09} (it is for example measured in nuclear scattering experiments).
\end{itemize}
%

Instead of the traditional DFT potential $\sum_{i=1}^N v_{ext}(\mathbf{r}_i)$, one might be tempted to introduce an arbitrary translational invariant potential of the form $\sum_{i=1}^N v_{int}(\mathbf{r}_i - \mathbf{R})$, where $\vec{R}=\frac{1}{N}\sum_{j=1}^{N}\vec{r}_j$ is the total c.m.\ of the particles.
This potential is an "internal" potential, i.e. is seen in the c.m. frame,
and in \cite{Mes09} we underlined that it is the only form which satisfies all the key formal properties.
However, $v_{int}$ should be zero in the purely isolated self-bound case.
This is why in \cite{Mes09} we presented it as a mathematical "auxiliary"
to reach our goal and showed that it can be dropped properly at the end, conserving all the conclusions.
Through it (and using the Jacobi coordinates), we proved, by a different way than those found in \cite{Eng07,Bar07},
the stationary "Internal DFT" theorem:
the internal many-body state can be written as a functional of $\rho_{int}$.
Then we formulated rigorously the corresponding "Internal" KS scheme (in the c.m.\ frame).
The main interest of this work is
to give a first step towards a fundamental justification to the use of internal density functionals
for stationary mean-field like calculations of nuclei \cite{Ben03} or He droplets \cite{Bar06} with effective interactions,
showing that there exists an ultimate functional which permits to reproduce the exact internal density, which was not clear up to now.

It is to be noted that the stationary Internal DFT / KS formalism 
gives a more fundamental justification than the Hartree-Fock (HF) framework
to the stationary nuclear mean-field like calculations. Indeed, HF does not contain quantum correlations,
nor treats correctly the redundant coordinate problem, which introduces a spurious coupling between the
internal properties and the c.m. motion \cite{RS80,Schm01a}.
A way to overcome this problem in the stationary case is to perform projected HF
(projection before variation on c.m.\ momentum), which permits to restore 
Galilean invariance, but at the price of abandoning the independent-particle 
description \cite{Schm01a,Pei62,Ben03}.
Within the Internal DFT / KS formalism,
we proved that the c.m.\ correlations can be included in the energy functional / the KS potential \cite{Mes09},
so that there would be no need for a c.m.\ projection if the ultimate functional was known.

It is a question of interest to generalize the stationary Internal DFT / KS formalism to the time-dependent case.
It would provide a first step towards a fundamental justification to the use of density functionals in nuclear
time-dependent calculations with an effective mean-field \cite{tdhf_nucl,Neg82},
and would prove that the c.m.\ correlations can be included in the functional.
This last point is even more interesting that the spurious c.m.\ motion problem remains in time-dependent HF \cite{Irv80,Uma09},
but that then the projected HF method becomes unmanageable and is not used in practise \cite{Uma09}.

In this paper, we propose to set up the time-dependent Internal DFT / KS formalism.
The paper is organized as follows:
we first apply the Jacobi coordinates method to the
time-dependent full many-body Hamiltonian to decouple the internal properties from the c.m. ones, and define some useful ``internal'' observables, including the internal density (section II);
then we show that the internal many-body wave function (and thus the ``internal'' mean values of all the observables) can be written as a functional of the internal density (section III);
finally, we develop the associated time-dependent Internal KS scheme as a practical scheme to compute the internal density (section IV).

\section{Time-dependent N body formulation.}

\subsection{General formulation.}

In the time-dependent domain, the introduction of an \textit{explicitely} time-dependent internal potential of the form
\begin{eqnarray}
\label{eq:v}
\sum_{i=1}^N v_{int}(\mathbf{r}_i - \mathbf{R};t)
\end{eqnarray}
takes a true meaning.
This is because self-bound systems are plagued by a c.m. problem.
For instance, in the stationary case, the c.m. will be delocalized in the whole space for \textit{isolated} self-bound systems \cite{Eng07,Kre01,Mes09}.
This does not occur in experiments because experimentally observed self-bound systems are not \textit{isolated} anymore
(they interact with the piece of matter they are inserted in which localizes the c.m.).
In the time domain, the c.m. motion remains uncomparable to the experimental one (this will be discussed in more detail later),
so that it would not make sense to introduce a time-dependent potential
which would act on the c.m. motion.
It are the internal properties which are of true experimental interest (experimentalists always deduce those properties \cite{foot4}).
This justifies
the introduction of an \textit{explicitely} time-dependent potential 
of the form (\ref{eq:v}),
which would act on the internal properties only,
and models the internal effect (only) of time-dependent potentials used in experiments.
Such a potential does not appear any more simply as a mathematical auxiliary (as for the stationary Internal DFT / KS) and should not necessarily be dropped at the end.


We thus start from a general translationally invariant $N$-body Hamiltonian
composed of the usual kinetic energy term, a 
translationally invariant 
two-body potential $u$, which describes the particle-particle 
interaction,
and an arbitrary translationally invariant "internal" potential $v_{int}$ which contains an explicit time dependence
%
\begin{equation}
\label{eq:H}
H
=   \sum_{i=1}^{N} \frac{\vec{p}^2_i}{2m} 
  + \sum_{\stackrel{i,j=1}{i > j}}^{N} u (\vec{r}_i-\vec{r}_j) 
  + \sum_{i=1}^{N} v_{\text{int}} (\vec{r}_i - \vec{R} ; t)
\; .
\end{equation}
For the sake of simplicity we assume a 2-body interaction $u$ and $N$ identical Fermions or Bosons.
The generalization to 3-body etc interactions is straightforward;
the generalization to different types of particles is underway.

We rewrite the Hamiltonian (\ref{eq:H}) using the ($N-1$) Jacobi coordinates $\{\xi_\alpha;\alpha=1,\dots,N-1\}$
and the c.m.\ coordinate $\vec{R}$, defined as
\begin{eqnarray}
&& \mathbf{\xi}_{1} = \mathbf{r}_2-\vec{r}_1, 
\mathbf{\xi}_2=\mathbf{r}_3-\frac{\vec{r}_2+\vec{r}_1}{2}, \ldots,
\nonumber\\
&& \mathbf{\xi}_{N-1} = \frac{N}{N-1} \, (\vec{r}_N - \vec{R}),
\nonumber\\
&& \vec{R}=\frac{1}{N}\sum_{j=1}^{N}\vec{r}_j
.
\label{eq:jacobi}
\end{eqnarray}
The $\xi_\alpha$ are relative to the c.m.\ of the other 
$1, \ldots, \alpha-1$ particles and are independent from $\vec{R}$. They are to be distinguished from the $N$ 
"laboratory coordinates" $\vec{r}_i$, and the $N$ "c.m. frame coordinates" $(\vec{r}_i-\vec{R})$ relative to the total c.m. $\vec{R}$.
As the $\{\vec{r}_i-\vec{r}_{j\ne i}\}$ and the $\{\vec{r}_i - \vec{R}\}$ can be rewritten as functions of the $\xi_\alpha$
(in Appendix \ref{app:jacobi} is given the expression of the $\{\mathbf{r}_i-\mathbf{R}\}$ as a function of the $\{\xi_\alpha\}$ coordinates),
the interaction $u$ and the internal potential $v_{int}$
can be rewritten as functions of the $\xi_\alpha$. We denote $U$ and $V$
the interaction potential and the internal potential in the Jacobi coordinates representation:
\begin{eqnarray}
\sum_{\stackrel{i,j=1}{i > j}}^{N} u (\vec{r}_i-\vec{r}_j)
\quad&\rightarrow&\quad
U(\mathbf{\xi}_1, ..., \mathbf{\xi}_{N-1}) 
\nonumber\\
\sum_{i=1}^{N} v_{int} (\vec{r}_i - \vec{R} ; t)
\quad&\rightarrow&\quad
V(\mathbf{\xi}_1, ..., \mathbf{\xi}_{N-1};t)
\label{eq:V_int}
.
\end{eqnarray}
Of course we have $U[u]$ and $V[v_{int}]$.
%
The $V[v_{int}]$ potential is ($N-1$) body in the Jacobi coordinates representation and
\textit{cannot} be written in a simple form in this representation (see Appendix \ref{app:jacobi}).
Moreover, various $v_{int}$ can lead to the same $V$, which we will develop later.

After having defined the conjugate momenta of $\vec{R}$ and $\xi_\alpha$,
we can separate (\ref{eq:H}) into $H = H_\text{CM} + H_\text{int}$, where ($M = Nm$ is the total mass)
\begin{equation}
H_\text{CM} = -\frac{\hbar^2 \Delta_\vec{R}}{2M} 
\label{eq:H_cm}
\end{equation}
is a one-body operator acting in  $\vec{R}$ space only,
and ($\tau_\alpha$ is the conjugate momentum of $\xi_\alpha$ and $\mu_\alpha = m\frac{\alpha}{\alpha+1}$ the corresponding reduced mass)
\begin{eqnarray}
H_\text{int}=\sum_{\alpha=1}^{N-1} \frac{\tau_\alpha^2}{2\mu_\alpha} &+& U[u](\mathbf{\xi}_1, ..., \mathbf{\xi}_{N-1}) 
\nonumber\\
&+& V[v_{int}](\mathbf{\xi}_1, ..., \mathbf{\xi}_{N-1};t)
\label{eq:H_int}
\end{eqnarray}
is a $(N-1)$ body operator in the $\{\xi_\alpha\}$ space. It contains the 
interaction and the internal potential.

In the time-dependent case, we can choose freely the initial state $\psi(\vec{r}_1, \ldots , \vec{r}_N ; t_0)$.
We start from an initial state which can be written 
\begin{equation}
\label{eq:psi_init}
\psi(\vec{r}_1, \ldots , \vec{r}_N ; t_0) 
= \Gamma(\vec{R} ; t_0) \; 
  \psi_{int} ({\boldmath{\xi}}_1, \ldots , {\boldmath{\xi}}_{N-1} ; t_0)
\end{equation}
in the Jacobi coordinates representation.
This form does not mix the c.m. motion with the internal one
(mixing them would not make sense because the c.m. motion does anyway not correspond to the experimental one)
and corresponds to the form of the stationary state  \cite{Schm01a,Mes09}.
As $H_{CM}$ and $H_{int}$ act in two separate subspaces, the $\mathbf{R}$ and $\{\xi_\alpha\}$ spaces
(which implies $[H_{CM},H_{int}]=0$), it is easy to show that the state $|\psi(t))$ can be built at all time $t\geq t_0$ as a direct product of the form
\begin{equation}
\label{eq:psi}
\psi(\vec{r}_1, \ldots , \vec{r}_N ; t) 
= \Gamma(\vec{R} ; t) \; 
  \psi_{int} ({\boldmath{\xi}}_1, \ldots , {\boldmath{\xi}}_{N-1} ; t) ,
\end{equation}
with
\begin{eqnarray}
&& H_{CM}|\Gamma(t)) = i\hbar\partial_t |\Gamma(t))
\label{eq:schro_int}\\
&& H_{int}|\psi_{int}(t)) = i\hbar\partial_t |\psi_{int}(t))
\label{eq:schro}
\; .
\end{eqnarray}
Hence, the $N$-body wave function $\psi$ can be separated into a one-body wave function 
$\Gamma$ that depends on the position $\mathbf{R}$ of the c.m. only, 
and an "internal" ($N-1$) body wave function 
$\psi_{int}$ that depends on the remaining ($N-1$) Jacobi 
coordinates ${\boldmath{\xi}}_\alpha$.
Of course, $\psi_{int}$ could also be written as a function of the $N$ laboratory coordinates $\mathbf{r}_i$,
but one of them would be redundant.
$\Gamma$ is solution of the free Schr\"odinger equation and describes the motion of the \textit{isolated} system as 
a whole in any chosen inertial frame of reference (as the laboratory).
If one starts from a normalizable initial state $|\Gamma(t_0))$, $|\Gamma(t))$
is condemned to spread more and more.
In the stationary limit, the only solutions of Eq. (\ref{eq:schro_int}) are plane waves, which are infinitely spread (thus not normalizable).
This does not correspond to experimental situations, where the system is not isolated anymore: interactions with other systems
of the experimental apparatus localize the c.m.
%
%
But the formal decoupling between the c.m.\ motion and the internal properties obtained when using the Jacobi coordinates method
permits to let the c.m.\ motion to the choice of experimental conditions,
the internal properties being comparable to the experimental ones.
%

\subsection{Some useful definitions.}
\label{par:def}

We define some quantities and relations that will be useful for the next considerations.
In \cite{Mes09,Gir08b,Kaz86} is defined the internal one-body density
\begin{eqnarray}
\lefteqn{\rho_{int}(\vec{r},t)/N}
\label{eq:rho_int0}
\\
& = & \int \! d\vec{r}_1 \cdots d\vec{r}_N \;
      \delta(\mathbf{R})
      |\psi_{int}(\vec{r}_1, \ldots, \vec{r}_{N};t)|^2
      \delta \big( \vec{r} - (\vec{r}_i-\mathbf{R}) \big)\, 
\nonumber\\
& = & \Big(\frac{N}{N-1}\Big)^3 
      \int \! d\vec{\xi}_1 \cdots  d\mathbf{\xi}_{N-2} \; 
      \big| \psi_{int} \big(\mathbf{\xi}_1, \ldots, \mathbf{\xi}_{N-2},
                   \tfrac{N\vec{r}}{N-1} ;t \big) \big|^2
\nonumber .
\end{eqnarray}
It is is normalized to $N$.
The laboratory density $\rho(\mathbf{r},t)$ is obtained by convolution of $\rho_{int}$ with the c.m.\ wave 
function (following \cite{Gir08b,Kaz86}):
$
\rho(\mathbf{r},t) = \int d\mathbf{R} |\Gamma(\mathbf{R},t)|^2 \rho_{int}(\mathbf{r} - \mathbf{R},t) .
$

We also introduced in \cite{Mes09} the local part of the two-body internal density matrix
\begin{eqnarray}
\label{eq:gamint0}
\lefteqn{\gamma_{int}(\vec{r},\vec{r'};t)}
      \\
& = & \int \! d\vec{r}_1 \cdots d\vec{r}_N \; 
      \delta(\mathbf{R}) |\psi_{int}(\vec{r}_1, \ldots, \vec{r}_{N};t)|^2 \,
      \nonumber \\
&   & \hspace{1.cm} \times 
      \delta \big( \vec{r} - (\vec{r}_i-\mathbf{R}) \big)
      \delta \big( \vec{r'} - (\vec{r}_{j\ne i}-\mathbf{R}) \big)
      \nonumber \\
& = & \frac{N(N-1)}{2} \Big(\frac{N-1}{N-2}\Big)^3 \Big(\frac{N}{N-1}\Big)^3
      \int \! d\mathbf{\xi}_1 \cdots  d\mathbf{\xi}_{N-3}
      \nonumber\\
&   & \hspace{1.cm} \times \Big| \psi_{int} \Big(\mathbf{\xi}_1, \ldots, \mathbf{\xi}_{N-3}, 
      \tfrac{\vec{r'}+(N-1)\vec{r} }{N-2},\tfrac{N\vec{r'}}{N-1} ;t \Big) \Big|^2
      \nonumber
.
\end{eqnarray}
It has the required normalisation to $N(N-1)/2$.
Following similar steps than in \cite{Gir08b,Kaz86},
we can show that the local part of the two-body laboratory density matrix 
$\gamma(\vec{r},\vec{r'},t)$ is obtained by convolution of $\gamma_{int}$ with the c.m.\ wave 
function:
$
\gamma(\vec{r},\vec{r'};t) = \int d\mathbf{R} |\Gamma(\mathbf{R},t)|^2 \gamma_{int}(\vec{r} - \mathbf{R},\vec{r'} - \mathbf{R};t) .
$

The definitions of $\rho_{int}(\vec{r},t)$ and $\gamma_{int}(\vec{r},\vec{r'};t)$ show clearly that they are defined in the c.m. frame,
i.e.\ that the $\vec{r}$, $\vec{r'}$ coordinates are measured in the c.m. frame (see the $\delta$ relations in (\ref{eq:rho_int0}) and (\ref{eq:gamint0})).
Compared to the traditional definitions, a $\delta(\mathbf{R})$ appears in the definition of the internal densities calculated
with $\psi_{int}$ in $\{\mathbf{r}_i\}$ coordinates. As one of them is redundant, the $\delta(\mathbf{R})$ represents the dependence of the redundant coordinate on the others \cite{foot3}.

Another quantity that will be very useful is the one-body internal probability current, defined in Appendix \ref{app:int_current} ($c.c.$ denotes the complex conjugate)
\begin{eqnarray}
\lefteqn{\mathbf{j}_{int}(\mathbf{r},t)/N}
\label{eq:j_int1}\\
&=& \frac{\hbar}{2m i} \Big(\frac{N}{N-1}\Big)^3 \int d\mathbf{\xi}_1 ...  d\mathbf{\xi}_{N-2} \psi_{int}^*(\mathbf{\xi}_1, ..., \mathbf{\xi}_{N-2}, \nu;t)
\nonumber\\
&& \hspace{1.5cm}
\times \mathbf{\nabla_{\nu}} \psi_{int}(\mathbf{\xi}_1, ..., \mathbf{\xi}_{N-2}, \nu;t) \Big|_{\nu=\frac{N}{N-1}\mathbf{r}} + c.c.
\nonumber
\end{eqnarray}
which satisfies the ``internal'' continuity equation
\begin{eqnarray}
\partial_t \rho_{int}(\mathbf{r},t) + \mathbf{\nabla} . \mathbf{j}_{int}(\mathbf{r},t) =0 .
\label{eq:cont_rel}
\end{eqnarray}
Using (\ref{eq:j_int1}), (\ref{eq:H_int}) and (\ref{eq:schro}), we obtain the relation
\begin{widetext}
\begin{eqnarray}
\lefteqn{i \frac{\partial}{\partial t} \mathbf{j}_{int}(\mathbf{r},t)}
\label{eq:partial_j_int}
\\
&=& \frac{N}{2m i} \Big(\frac{N}{N-1}\Big)^3 \int d\mathbf{\xi}_1 ...  d\mathbf{\xi}_{N-2} \Big\{ 
\mathbf{\nabla_{\nu}} \psi_{int}(\mathbf{\xi}_1, ..., \mathbf{\xi}_{N-2},\nu;t) i\hbar\partial_t \psi_{int}^*(\mathbf{\xi}_1, ..., \mathbf{\xi}_{N-2},\nu;t) \nonumber\\
&& \hspace{4.6cm} +
\mathbf{\nabla_{\nu}} \Big( i\hbar \partial_t\psi_{int}(\mathbf{\xi}_1, ..., \mathbf{\xi}_{N-2},\nu;t) \Big) \psi_{int}^*(\mathbf{\xi}_1, ..., \mathbf{\xi}_{N-2},\nu;t)  + c.c.
\Big\}\Big|_{\nu=\frac{N}{N-1}\mathbf{r}}
\nonumber\\
&=& \frac{N}{2m i} \Big(\frac{N}{N-1}\Big)^3 \int d\mathbf{\xi}_1 ...  d\mathbf{\xi}_{N-2} \Big\{ 
\mathbf{\nabla_{\nu}} \psi_{int}(\mathbf{\xi}_1, ..., \nu;t)  \frac{\hbar^2\Delta_{\nu}}{2\mu_{N-1}} \psi_{int}^*(\mathbf{\xi}_1, ..., \nu;t)
-  \psi_{int}^*(\mathbf{\xi}_1, ..., \nu;t) \mathbf{\nabla_{\nu}} \frac{\hbar^2\Delta_{\nu}}{2\mu_{N-1}} \psi_{int}(\mathbf{\xi}_1, ..., \nu;t)
\nonumber\\
&& \hspace{4.5cm} +
\psi^*_{int}(\mathbf{\xi}_1, ..., \nu;t) \mathbf{\nabla_{\nu}} \Big( U[u](\mathbf{\xi}_1, ..., \nu) + V[v_{int}](\mathbf{\xi}_1, ..., \nu;t) \Big) \psi_{int}(\mathbf{\xi}_1, ... ,\nu;t) 
+c.c.
\Big\} \Big|_{\nu=\frac{N}{N-1}\mathbf{r}},
\nonumber
\end{eqnarray}
\end{widetext}
which will be a key equation for the next considerations.

\section{Time-dependent Internal DFT theorem.}

\subsection{Preliminaries.}
\label{sub:int}

To prove the time-dependent Internal DFT theorem, we adapt the considerations of \cite{Run84,Gro94}
to the internal Schr\"odinger equation (\ref{eq:schro}).
The main differences lie in the definition of the corresponding internal density (\ref{eq:rho_int0}) and probability current (\ref{eq:j_int1}),
and in the fact that the potential $V[v_{int}](\mathbf{\xi}_1, ..., \mathbf{\xi}_{N-1};t)$
cannot be written as the sum of one-body potentials in the Jacobi coordinates representation
(which introduces some subtleties due to the c.m. correlations and will bring us to use the integral mean value theorem to reach our goal).

In what follows, we consider a given type of Fermions or Bosons, i.e. a given particle-particle interaction $u$.
Solving the ``internal'' Schr\"odinger equation (\ref{eq:schro}) for a fixed initial state $|\psi_{int}(t_0))$ and for various internal potentials $V[v_{int}](\mathbf{\xi}_1, ..., \mathbf{\xi}_{N-1};t)$ defines two maps \cite{Run84,Gro94}
\begin{eqnarray}
&& F: V[v_{int}](\mathbf{\xi}_1, ..., \mathbf{\xi}_{N-1};t)\rightarrow |\psi_{int}(t))
\nonumber\\
&& G: V[v_{int}](\mathbf{\xi}_1, ..., \mathbf{\xi}_{N-1};t)\rightarrow \rho_{int}(\mathbf{r},t).
\label{eq:map}
\end{eqnarray}
We first notice that two potentials $v_{int}$ and $v'_{int}$ which lead to two potentials
$V[v_{int}](\mathbf{\xi}_1, ..., \mathbf{\xi}_{N-1};t)$ and $V[v'_{int}](\mathbf{\xi}_1, ..., \mathbf{\xi}_{N-1};t)$ that differ by a scalar function of time only $c(t)$, 
will give two wave functions that differ by a phase $e^{-i\alpha(t)/\hbar}$ only \cite{Run84,Gro94}:
\begin{eqnarray}
&&V[v'_{int}](\mathbf{\xi}_1, ..., \mathbf{\xi}_{N-1};t) - V[v_{int}](\mathbf{\xi}_1, ..., \mathbf{\xi}_{N-1};t) = c(t)
\nonumber\\
&&\quad\quad\Rightarrow\hspace{1mm}
|\psi'_{int}(t))=e^{-i\alpha(t)/\hbar}|\psi_{int}(t)) ,
\nonumber\\
&& \hspace{2.5cm} \rm{with} \hspace{2mm} \Dot{\alpha}(t)=c(t)
\label{eq:v'}
.
\end{eqnarray}
Then, $|\psi_{int}(t))$ and $|\psi'_{int}(t))$ will give the same density $\rho_{int}(\mathbf{r},t)=\rho'_{int}(\mathbf{r},t)$.
The consequence is that the map $G$ is not fully invertible.

Let us discuss a bit about the condition (\ref{eq:v'}).
The form (\ref{eq:V_int}) for $V[v_{int}]$ implies $V[v'_{int}]-V[v_{int}]=V[v'_{int}-v_{int}]$.
We define
\begin{eqnarray}
\Delta v_{int}(\mathbf{r};t)=v'_{int}(\mathbf{r};t)-v_{int}(\mathbf{r};t).
\label{eq:delta_v}
\end{eqnarray}
It is to be noted that the condition $\Delta v_{int}(\mathbf{r};t)\ne c(t)/N$
is necessary but not sufficient to ensure the condition (\ref{eq:v'}),
which can be rewritten $V[\Delta v_{int}](\mathbf{\xi}_1, ..., \mathbf{\xi}_{N-1};t)\ne c(t)$.
Indeed, it is possible to have $\Delta v_{int}(\mathbf{r};t)\ne c(t)/N$ and nevertheless
$V[\Delta v_{int}](\mathbf{\xi}_1, ..., \mathbf{\xi}_{N-1};t) = c(t)$,
because compensations due to the c.m. correlations can happen.

Let us reason on the two particules case, where only one Jacobi coordinate is sufficient to describe the internal properties.
We have (see Appendix \ref{app:jacobi}):
$V[\Delta v_{int}](\mathbf{\xi}_1;t)=\Delta v_{int} (-\frac{1}{2}\xi_1;t)+\Delta v_{int} (\frac{1}{2}\xi_1;t) \big( =\sum_{i=1}^{2} v_{\text{int}} (\vec{r}_i - \vec{R} ; t) \big)$.
We see that if $\Delta v_{int}(\mathbf{r};t)$ is an odd function of $\mathbf{r}$ at all $t$
(up to an additional time-dependent function),
we have $V[\Delta v_{int}](\mathbf{\xi}_1;t)=c(t)$
$\Rightarrow$ $\rho_{int}=\rho'_{int}$.
This is due to the c.m. correlations, that the non-trivial form of $V$ reflects.
If
$\Delta v_{int}$ tends to move the first particule in one direction,
the second particule will tend to move in the opposite direction because of the c.m. correlations.
But if this potential counter-acts perfectly the motion of the second particule (as does an odd potential in the c.m. frame), then the particules remain stuck and the density unchanged.

The same can occur for an arbitrary number of particules.
For instance, as $\sum_{i=1}^N (\mathbf{r}_i-\mathbf{R})=0$,
it is obvious with (\ref{eq:V_int}) and (\ref{eq:delta_v}) that every $\Delta v_{int}(\mathbf{r};t)=\mathbf{b}(t).\mathbf{r}+c(t)/N$
will yield $V[\Delta v_{int}] = c(t)$ (even if this form for $\Delta v_{int}$ leads to internal potentials which are not null at infinity).
Again, this is because if a potential counter acts perfectly the motion due to the c.m. correlations, the particules remain stuck and the density unchanged.
In what follows, we consider only internal potentials $v_{int}$ and $v'_{int}$ that lead to $V[\Delta v_{int}]\ne c(t)$.

We come back to Eq. (\ref{eq:v'}) and denote $|\psi_{int}(t))=e^{-i\alpha(t)/\hbar}|\psi^0_{int}(t))$ where we define $\psi^0_{int}$ as the wave function obtained for the choice $c(t)=0$,
i.e. associated to a $V[v_{int}](\mathbf{\xi}_1, ..., \mathbf{\xi}_{N-1};t)$ where no additive time-dependent function can be split.
If we prove that the map $G$ is invertible up to an additive time-dependent function $c(t)$, then $\psi^0_{int}$ is fixed by $\rho_{int}$
through the relation $|\psi^0_{int}(t))=F G^{-1} \rho_{int}(\mathbf{r},t)$,
which implies that $|\psi^0_{int}(t))$ can be written as a functional of the internal density $\rho_{int}$ defined in (\ref{eq:rho_int0}).
Consequently, any expectation value of an operator $\hat{O}$ which does not contain a time derivative can be written as a functional of the internal density (as the phase cancels out):
$(\psi_{int}(t)|\hat{O}|\psi_{int}(t))=(\psi^0_{int}[\rho_{int}](t)|\hat{O}|\psi^0_{int}[\rho_{int}](t))$.

We thus have to show that 
a propagation of (\ref{eq:schro}) with two potentials $v_{int}$ and $v_{int}'$ that yield
$V[\Delta v_{int}](\mathbf{\xi}_1, ..., \mathbf{\xi}_{N-1};t)\ne c(t)$
will produce two different internal densities $\rho_{int}$ and $\rho_{int}'$.

\subsection{The proof.}

We start from a \textit{fixed initial state} $|\psi_{int}(t_0))$ and
propagate it with two with two potentials $v_{int}$ and $v_{int}'$ that give
$V[\Delta v_{int}](\mathbf{\xi}_1, ..., \mathbf{\xi}_{N-1};t)\ne c(t)$.
We deduce from Eq. (\ref{eq:partial_j_int})
\begin{widetext}
\begin{eqnarray}
i \frac{\partial}{\partial t} \Big( \mathbf{j}_{int}(\mathbf{r},t) - \mathbf{j}'_{int}(\mathbf{r},t) \Big) \Big|_{t=t_0}  =
\frac{N}{m i} \Big(\frac{N}{N-1}\Big)^3
\int d\mathbf{\xi}_1 ...  d\mathbf{\xi}_{N-2} |\psi_{int}(\mathbf{\xi}_1, ..., \nu;t_0)|^2
\mathbf{\nabla_{\nu}} V[\Delta v_{int}](\mathbf{\xi}_1, ..., \nu;t_0) \Big|_{\nu=\frac{N}{N-1}\mathbf{r}} .
\end{eqnarray}
Using the ``internal'' continuity relation (\ref{eq:cont_rel}) we obtain
\begin{eqnarray}
\lefteqn{\frac{\partial^2}{\partial t^2} \Big( \rho_{int}(\mathbf{r},t) - \rho'_{int}(\mathbf{r},t) \Big) \Big|_{t=t_0} =}
\label{eq:partial_j_int0}\\
&& \frac{N}{m} \Big(\frac{N}{N-1}\Big)^3 \mathbf{\nabla_{\mathbf{r}}} .\int d\mathbf{\xi}_1 ...  d\mathbf{\xi}_{N-2}
|\psi_{int}(\mathbf{\xi}_1, ..., \frac{N}{N-1}\mathbf{r};t_0)|^2
\mathbf{\nabla}_\nu V[\Delta v_{int}](\mathbf{\xi}_1, ..., \nu;t_0) \Big|_{\nu=\frac{N}{N-1}\mathbf{r}}
.
\nonumber
\end{eqnarray}

We now make the only hypothesis which is used in this derivation. Following \cite{Run84,Gro94} we restrict the set of
potentials $v_{int}$ to those that can be expanded into Taylor series with respect to the time at the initial time $t_0$
(which is a reasonable hypothesis for physical potentials).
As we supposed that $V[\Delta v_{int}](\mathbf{\xi}_1, ..., \mathbf{\xi}_{N-1};t)\ne c(t)$, we have ($k$ is a positive integer)
\begin{eqnarray}
V[\Delta v_{int}](\mathbf{\xi}_1, ..., \mathbf{\xi}_{N-1};t)\ne c(t)
\quad\Rightarrow\quad
\exists k :
w_k(\mathbf{\xi}_1, ..., \mathbf{\xi}_{N-1};t_0) \ne constant ,
\label{eq:Vint}
\end{eqnarray}
where
\begin{eqnarray}
w_k(\mathbf{\xi}_1, ..., \mathbf{\xi}_{N-1};t_0)
= \frac{\partial^k}{\partial t^k} V[\Delta v_{int}](\mathbf{\xi}_1, ..., \mathbf{\xi}_{N-1};t) \Big|_{t=t_0}
\label{eq:wk}
.
\end{eqnarray}
%
%
%
It is to be noted that the condition $\frac{\partial^k}{\partial t^k} \Delta v_{int}(\mathbf{r};t)\Big|_{t=t_0} \ne constant$
$\Rightarrow$ $\mathbf{\nabla_r} \frac{\partial^k}{\partial t^k} \Delta v_{int}(\mathbf{r};t)\Big|_{t=t_0}\ne \overrightarrow{0}$
is necessary to ensure the condition (\ref{eq:Vint}) (see (\ref{eq:V_int}) and (\ref{eq:wk})), but not sufficient.

In what follows, we consider $k$ as the \textit{smallest} positive integer such that (\ref{eq:Vint})
is verified. Then, if we apply $k$ time derivatives to the Eq. (\ref{eq:partial_j_int0}), we straightforwardly obtain
\begin{eqnarray}
\frac{\partial^{k+2}}{\partial t^{k+2}} \Big( \rho_{int}(\mathbf{r},t) - \rho_{int}'(\mathbf{r},t) \Big) \Big|_{t=t_0} =
\frac{N}{m} \Big(\frac{N}{N-1}\Big)^3 \mathbf{\nabla_{\mathbf{r}}} . \int d\mathbf{\xi}_1 ...  d\mathbf{\xi}_{N-2} |\psi_{int}(\mathbf{\xi}_1, ..., \nu;t_0)|^2 \mathbf{\nabla_\nu} w_k(\mathbf{\xi}_1, ..., \nu;t_0) \Big) \Big|_{\nu=\frac{N}{N-1}\mathbf{r}}
.
\end{eqnarray}
As, for every physical potential, $\mathbf{\nabla}_{\mathbf{\xi}_{N-1}} w_k(\mathbf{\xi}_1, ..., \mathbf{\xi}_{N-1};t_0)$ is a real and continuous function in the whole position space, and as $|\psi_{int}(\mathbf{\xi}_1, ...,\mathbf{\xi}_{N-1};t_0)|^2$ is a real and positive function in the whole position space, we can apply the integral mean value theorem generalized to many variables functions (demonstrated in Appendix \ref{app:mean_val_th}) to the previous expression. We obtain
\begin{eqnarray}
\exists (\beta_1,...,\beta_{N-2}) \hspace{1mm} :
&& m \frac{\partial^{k+2}}{\partial t^{k+2}} \Big( \rho_{int}(\mathbf{r},t) - \rho_{int}'(\mathbf{r},t) \Big) \Big|_{t=t_0}
\nonumber\\
&& \hspace{5mm} = \mathbf{\nabla_{\mathbf{r}}} . \Big[
\mathbf{\nabla}_\frac{N\mathbf{r}}{N-1} w_k(\beta_1,...,\beta_{N-2}, \frac{N}{N-1}\mathbf{r};t_0)
N \Big(\frac{N}{N-1}\Big)^3 \int d\mathbf{\xi}_1 ...  d\mathbf{\xi}_{N-2}|\psi_{int}(\mathbf{\xi}_1, ..., \frac{N}{N-1}\mathbf{r};t_0)|^2 \Big]
\nonumber\\
&& \hspace{5mm} = \mathbf{\nabla_{\mathbf{r}}} . \Big[
\mathbf{\nabla}_\frac{N\mathbf{r}}{N-1} w_k(\beta_1,...,\beta_{N-2}, \frac{N}{N-1}\mathbf{r};t_0)
\rho_{int}(\mathbf{r},t_0) \Big]
.
\label{eq:partial_j_int2}
\end{eqnarray}
%
%
%
%
To prove the one-to-one correspondence $V[v_{int}](\mathbf{\xi}_1, ..., \mathbf{\xi}_{N-1};t)\leftrightarrow \rho_{int}(\mathbf{r},t)$ it remains to show that (\ref{eq:partial_j_int2})
cannot vanish for $v_{int}$ and $v_{int}'$ that lead to the relation (\ref{eq:Vint}).
Then the internal densities $\rho_{int}(\mathbf{r},t)$ and $\rho_{int}'(\mathbf{r},t)$ would become different infinitesimally later than $t_0$.
We use the \textit{reductio ad absurdum} method, in the spirit of Refs. \cite{Run84,Gro94}. We suppose that (\ref{eq:partial_j_int2}) vanishes, which implies:
\begin{eqnarray}
%
0&=&\frac{N-1}{N}\int d\mathbf{r} w_k(\beta_1,...,\beta_{N-2}, \frac{N}{N-1}\mathbf{r};t_0) \mathbf{\nabla_{\mathbf{r}}} . \Big[
\mathbf{\nabla}_\frac{N\mathbf{r}}{N-1} w_k(\beta_1,...,\beta_{N-2}, \frac{N}{N-1}\mathbf{r};t_0)
\rho_{int}(\mathbf{r},t_0) \Big]
\nonumber\\
&=& 
-\int d\mathbf{r} \Big[ \mathbf{\nabla}_\frac{N\mathbf{r}}{N-1} w_k(\beta_1,...,\beta_{N-2}, \frac{N}{N-1}\mathbf{r};t_0) \Big]^2
\rho_{int}(\mathbf{r},t_0)
.
\label{eq:absurdum2}
\end{eqnarray}
As $w_k$ is a many-body function, the Eq. (\ref{eq:Vint}) does not imply that
$\forall (\beta_1,...,\beta_{N-2}):\mathbf{\nabla}_{\xi_{N-1}} w_k(\beta_1,...,\beta_{N-2}, \xi_{N-1};t_0)\ne\overrightarrow{0}$ in the general case.
However, we check if this relation holds for the particular form (\ref{eq:V_int}) we choose for $V$.

Inserting the results of Appendix \ref{app:jacobi} in (\ref{eq:V_int}) and (\ref{eq:wk}), we obtain, if $N>2$ 
(the case $N=2$ will be discussed later on)
\begin{eqnarray}
&& w_k(\beta_1,...,\beta_{N-2}, \xi_{N-1};t_0)=
\label{eq:w_k}\\
&&\quad
\frac{\partial^{k}}{\partial t^{k}} \Delta v_{int} \big(\frac{N-1}{N}\xi_{N-1};t\big) \Big|_{t=t_0}
+ \sum_{i=1}^{N-2}\frac{\partial^{k}}{\partial t^{k}} \Delta v_{int} \big(\gamma_i-\frac{1}{N}\xi_{N-1};t\big) \Big|_{t=t_0}
+\frac{\partial^{k}}{\partial t^{k}} \Delta v_{int} \big(-\sum_{i=1}^{N-2}\gamma_i-\frac{1}{N}\xi_{N-1};t\big) \Big|_{t=t_0} ,
\nonumber
\end{eqnarray}
where we defined
\begin{eqnarray}
\gamma_{N-2}=\frac{N-2}{N-1}\beta_{N-2}
\quad\quad \text{and} \quad\quad
\forall i \in [1,N-3]: \gamma_{i}=\frac{i}{i+1}\beta_{i} - \sum_{\alpha=i+1}^{N-2} \frac{1}{\alpha+1}\beta_{\alpha}.
\label{eq:gamma}
\end{eqnarray}
The form of the third term of the right hand side of Eq. (\ref{eq:w_k}) comes from the fact that $\sum_{i=1}^N(\mathbf{r}_i-\mathbf{R})=0$, which implies, using the Appendix \ref{app:jacobi}, that $-\sum_{\alpha=1}^{N-2} \frac{1}{\alpha+1}\beta_{\alpha}=-\sum_{i=1}^{N-2}\gamma_i$. 
We see from Eq. (\ref{eq:gamma}) that the set $(\gamma_1,...,\gamma_{N-2})$ is perfectly defined by the set $(\beta_1,...,\beta_{N-2})$ and vice versa.
We now can calculate
\begin{eqnarray}
&&\mathbf{\nabla}_{\xi_{N-1}} w_k(\beta_1,...,\beta_{N-2}, \xi_{N-1};t_0)=
\nonumber\\
&&\hspace{3cm}
\frac{N-1}{N} \mathbf{D} \big(\frac{N-1}{N}\xi_{N-1}\big)-\frac{1}{N}\sum_{i=1}^{N-2} \mathbf{D} \big(\gamma_i-\frac{1}{N}\xi_{N-1}\big)
-\frac{1}{N} \mathbf{D} \big(-\sum_{i=1}^{N-2}\gamma_i-\frac{1}{N}\xi_{N-1}\big)
\label{eq:w}
,
\end{eqnarray}
where we introduced
\begin{eqnarray}
\mathbf{D}(\mathbf{r})=\mathbf{\nabla}_\mathbf{r} \frac{\partial^{k}}{\partial t^{k}} \Delta v_{int} (\mathbf{r};t) \big|_{t=t_0}
.
\label{eq:D}
\end{eqnarray}
for simplicity.
%
%
We now check if
$\exists (\beta_1,...,\beta_{N-2}):$ $\mathbf{\nabla}_{\xi_{N-1}} w_k(\beta_1,...,\beta_{N-2}, \xi_{N-1};t_0)=\overrightarrow{0}$,
which is equivalent, according to (\ref{eq:gamma}) and (\ref{eq:w}), to check if
\begin{eqnarray}
\exists (\gamma_1,...,\gamma_{N-2}):\quad
(N-1) \mathbf{D} \big((N-1)\mathbf{r}\big)=\sum_{i=1}^{N-2} \mathbf{D} \big(\gamma_i-\mathbf{r}\big)
+\mathbf{D} \big(-\sum_{i=1}^{N-2}\gamma_i-\mathbf{r}\big)
.
\end{eqnarray}
\end{widetext}
%
%
Some mathematical considerations show that this equation cannot be fulfilled for all $\mathbf{r}$ when $N>2$, whatever the set of $(\gamma_1,...,\gamma_{N-2})$,
instead if $\mathbf{D} (\mathbf{r})=\overrightarrow{const}$.
But if $\mathbf{D} (\mathbf{r})=\overrightarrow{const.}$, then $\Delta v_{int} (\mathbf{r};t)$ should be 
equal to $\mathbf{b}(t).\mathbf{r}+c(t)/N$, according to (\ref{eq:D}),
which is forbidden by the condition (\ref{eq:Vint}), cf. discussion of the \S \ref{sub:int}.

It remains to discuss the case $N=2$. It is easy to show that then, we have
$\mathbf{\nabla}_{\xi_{N-1}} w_k(\xi_{N-1};t_0)=
-\frac{1}{2}\mathbf{D}\big(-\frac{1}{2}\xi_{N-1}\big)+\frac{1}{2}\mathbf{D}\big(\frac{1}{2}\xi_{N-1}\big)$,
which is null if $\mathbf{D}(\mathbf{r})$
is any par function of $\mathbf{r}$.
But if $\mathbf{D} (\mathbf{r})$ is par, then $\frac{\partial^{k}}{\partial t^{k}}\Delta v_{int} (\mathbf{r};t)$ should be an odd function of $\mathbf{r}$
(up to an additional time-dependnt function), according to (\ref{eq:D}),
which is also forbidden by the condition (\ref{eq:Vint}), cf. discussion of the \S \ref{sub:int}.

Thus, we can conclude that, in our case
\begin{eqnarray}
\forall (\beta_1,...,\beta_{N-2}):
\mathbf{\nabla}_{\xi_{N-1}} w_k(\beta_1,...,\beta_{N-2}, \xi_{N-1};t_0)\ne\overrightarrow{0}
.
\nonumber
\label{eq:absurdum1}
\end{eqnarray}
%
%
%
We immediately deduce the incompatibility of
this relation,
which is a consequence of (\ref{eq:Vint}) and of the particular form (\ref{eq:V_int}) of $V$, with (\ref{eq:absurdum2}). Thus, the hypothesis we made is absurd: Eq. (\ref{eq:partial_j_int2}) cannot vanish if $V[\Delta v_{int}]\ne c(t)$, so that the internal densities $\rho_{int}(\mathbf{r},t)$ and $\rho_{int}'(\mathbf{r},t)$ become different infinitesimally later than $t_0$. As a consequence, the map $G$, defined in (\ref{eq:map}), is invertible (up to an additive time-dependent function) and $|\psi^0_{int}(t))$ can be written as a functional of the internal density (we use the notation (\ref{eq:v'})).
Thus, any expectation value of an operator $\hat{O}$ which does not contain a time derivative can be written as a functional of $\rho_{int}$ as the phase cancels out.
This achieves to prove the time-dependent Internal DFT theorem
(which is a variant of the Runge-Gross theorem \cite{Run84,Gro94} for self-bound systems and internal densities).

Mind that all the previous reasonings hold only for a fixed initial state $\psi_{int}(t_0)$ (and a given type of particle), so that $\psi^0_{int}$ is not only a functional of $\rho_{int}$, but also depends on $\psi_{int}(t_0)$. This will be discussed further.

\subsection{Link with traditional (time-dependent) DFT.}

We stress here the link and differences between the traditional DFT and internal DFT potentials.
We recall that the form of the potential $v_{ext}$ of traditional DFT can be fundamentally justified starting from the
laboratory Hamiltonian of
an isolated molecule where the nuclei are treated explicitely.
As a molecule is a self-bound system, one can apply the Jacobi coordinates method.
We denote the N electronic coordinates related to 
the laboratory frame
as $\mathbf{r}_i$, 
the nuclear c.m. coordinate as $\mathbf{R}^{nucl}$ and
the N electronic coordinates related to the c.m. of the nuclei as $\mathbf{r}'_i=\mathbf{r}_i - \mathbf{R}^{nucl}$.
A key point concerning the molecules is that, as the nuclei are much heavier than the electrons,
the c.m. of the whole molecule coincides with $\mathbf{R}^{nucl}$,
and it is an excellent approximation to apply the Jacobi coordinates to the nuclear coordinates only.
As a result, the c.m. motion will be described by a $\Gamma(\mathbf{R}^{nucl})$ wave function.
The redundant coordinate problem (thus the c.m.\ correlations) will concern the nuclei only, and will be ``external'' to the electronic problem:
the N electrons are still described by N coordinates.
Then, if one decouples the electronic motion from the nuclear one doing the clamped nuclei approximation,
the interaction of the electrons with the nuclear background is described by a potential of the form $\sum_{i=1}^N v_{ext}(\mathbf{r}_i - \mathbf{R}^{nucl})$,
which becomes $\sum_i v_{ext}(\mathbf{r}'_i)$ when moving to the c.m. frame.
We then recover the form of the traditional DFT potential.
The potential $v_{ext}$, which is \textit{internal} for the (self-bound) molecular problem, becomes \textit{external} for the pure electronic problem.
Those considerations also hold in the time domain, the difference being that the potential
\begin{eqnarray}
\sum_{i=1}^N v_{ext}(\mathbf{r}_i - \mathbf{R}^{nucl};t)
\label{eq:v_ext}
\end{eqnarray}
can then contain an explicit time dependence in addition to the part which
describes the interaction of the electrons with the nuclear background.
We recover the traditional time-dependent DFT potential \cite{Run84,Gro94,Gro90,Mar04} when moving in the c.m. frame.

Those reasonings explicit the link between the traditional DFT potential expressed with the laboratory coordinates, Eq. (\ref{eq:v_ext}),
and the Internal DFT potential expressed with the laboratory coordinates, Eq. (\ref{eq:v}).
They both act only on the internal properties, and not on the c.m. motion
(because it is anyway not comparable to the experimental one).
The difference is that as, in the molecular case, some particules are much heavier than the other,
it is a very good approximation to assimilate the c.m. of the whole molecule with $\mathbf{R}^{nucl}$,
which permits to neglect the c.m. correlations for the electronic system,
and to justify the clamped nuclei approximation.
This simplifies greatly the electronic problem and the traditional DFT can be used to study it.
When the particules constituting the self-bound system have nearly the same masses, as it is the case for the nuclei or the He droplets, 
the total c.m. ($\mathbf{R}$) should be calculated with \textit{all} the particules,
so that the c.m. correlations will concern all the particules, and no clamped approximation can be justified.
Then, we should use the formalism proposed here.

\section{Time-dependent Internal Kohn-Sham scheme.}

We now provide a practical scheme to calculate the internal density $\rho_{int}$, which consists in the generalization of the stationary Internal KS scheme of \cite{Mes09} to the time-dependent case.
First, we note that for any normalizable initial state $|\psi_{int}(t_0))$, which are the only allowed, the ``internal'' Schr\"odinger equation (\ref{eq:schro}) stems
from a variational principle
on the ``internal'' quantum action \cite{ker76,Run84,Vig08}
\begin{eqnarray}
A_{int} = \int_{t_0}^{t_1} dt (\psi_{int}(t)|i\hbar\partial_t-H_{int}|\psi_{int}(t)) .
\label{eq:action}
\end{eqnarray}
As the function $c(t)$ possibly contained in the potential $V_{int}$ is perfectly canceled by the time derivative of the corresponding phase $e^{-i\alpha(t)/\hbar}$ of $\psi_{int}$, see (\ref{eq:v'}),
we have $A_{int}=\int_{t_0}^{t_1} dt (\psi_{int}^0[\rho_{int}](t)|i\hbar\partial_t-H_{int}|\psi_{int}^0[\rho_{int}](t))$
if $V_{int}$ is chosen so that no additive time-dependent function can be split.
Thus, the internal quantum action can be considered as a functional of $\rho_{int}$.
Its
$\int_{t_0}^{t_1} dt (\psi_{int}^0(t)|i\hbar\partial_t-\sum_{\alpha=1}^{N-1} \frac{\tau_\alpha^2}{2\mu_\alpha} - U[u]|\psi_{int}^0(t))$
part is a universal functional of $\rho_{int}$ in the sense that, for a given type of particle (a given interaction $u$), the same dependence on $\rho_{int}$ holds for every $V[v_{int}]$, thus $v_{int}$ (see (\ref{eq:V_int})).

Using the Eq. (\ref{eq:H_int}), we develop the ``internal'' quantum action as
\begin{eqnarray}
A_{int}[\rho_{int}]&=& \int_{t_0}^{t_1} dt (\psi_{int}^0(t)|i\hbar\partial_t-\sum_{\alpha=1}^{N-1} \frac{\tau_\alpha^2}{2\mu_\alpha}|\psi_{int}^0(t))
\label{eq:action2}\\
&& - \int_{t_0}^{t_1} dt (\psi_{int}^0(t)|U[u](\mathbf{\xi}_1, ..., \mathbf{\xi}_{N-1})|\psi_{int}^0(t))
\nonumber\\
&& - \int_{t_0}^{t_1} dt (\psi_{int}^0(t)|V[v_{int}](\mathbf{\xi}_1, ..., \mathbf{\xi}_{N-1};t)|\psi_{int}^0(t))
\nonumber .
\end{eqnarray}
To rewrite its last two terms in a more convenient way, we establish a useful relation.
For any function $f(\vec{r}_1,...,\vec{r}_N;t)$ of the laboratory coordinates,
expressible with the Jacobi coordinates [we denote $F(\xi_1,...,\xi_{N-1};t)$], we have
\begin{eqnarray}
\label{eq:rel}
\lefteqn{(\psi_{int}^0(t)| F(\xi_1,...,\xi_{N-1};t) |\psi_{int}^0(t))}
\\
& = & \int \! d\mathbf{\xi}_1 \cdots  d\mathbf{\xi}_{N-1} F(\xi_1,...,\xi_{N-1};t) 
\big| \psi_{int}^0 (\xi_1,...,\xi_{N-1};t) \big|^2
\nonumber\\
& = & \int \! d\mathbf{R} d\mathbf{\xi}_1 \cdots  d\mathbf{\xi}_{N-1} \delta(\mathbf{R}) F(\xi_1,...,\xi_{N-1};t)
\nonumber\\
&& \hspace{4.5cm} \times \big| \psi_{int}^0 (\xi_1,...,\xi_{N-1};t) \big|^2
\nonumber\\
& = & \int \! d\vec{r}_1 \cdots d\vec{r}_{N} \delta(\mathbf{R}) f(\vec{r}_1,...,\vec{r}_N;t) \big| \psi_{int}^0(\vec{r}_1,...,\vec{r}_N;t) \big|^2
\nonumber
\, .
\end{eqnarray}
We see that the "internal mean values" calculated with $\psi_{int}$
expressed as a function of the ($N-1$) coordinates $\xi_\alpha$, can also be calculated with $\psi_{int}$
expressed as a function of the $N$ coordinates $\mathbf{r}_i$.
As one of them is redundant, a $\delta(\mathbf{R})$ which
represents the dependence of the redundant coordinate on the others appears \cite{foot3}.

The relation (\ref{eq:rel}) leads to
\begin{eqnarray}
\label{eq:Eext}
\lefteqn{
(\psi_{int}^0(t)|V[v_{int}](\xi_1,...,\xi_{N-1};t)|\psi_{int}^0(t)) 
} \nonumber\\
& = & \int \! d\vec{r}_1 \cdots d\vec{r}_N \; 
      \delta(\mathbf{R}) \sum_{i=1}^N v_{int}(\vec{r}_i - \vec{R};t) |\psi_{int}^0(\vec{r}_1,...,\vec{r}_N;t)|^2 \, 
      \nonumber\\
& = & \sum_{i=1}^N 
      \int \! d\vec{r} \; v_{int}(\vec{r};t) 
      \int \! d\vec{r}_1 \cdots d\vec{r}_N \; \delta(\mathbf{R})
      \nonumber\\
&   & \hspace{2cm} \times |\psi_{int}^0(\vec{r}_1, \ldots, \vec{r}_{N};t)|^2  \delta \big( \vec{r}-(\vec{r}_i-\vec{R}) \big) 
      \nonumber\\
& = & \sum_{i=1}^N 
      \int \! d\vec{r} \; v_{int}(\vec{r};t) \, 
      \, \frac{\rho_{int}(\vec{r},t)}{N} 
      \nonumber\\
& = & \int \! d\vec{r} \; v_{int}(\vec{r};t) \, \rho_{int}(\vec{r},t)
,
\end{eqnarray}
where we used (\ref{eq:rho_int0}) to obtain the penultimate equality.
We see that the potential $\sum_{i=1}^N v_{int}(\vec{r}_i - \vec{R};t)$ that is $N$ body 
with respect to the laboratory coordinates (and $(N-1)$ body when 
expressed with Jacobi coordinates), becomes one body (and local) when
expressed with the c.m. frame coordinates
(mind that $\rho_{int}$ is defined in the c.m. frame, i.e. that $\mathbf{r}$ is measured in the c.m. frame, cf. \S \ref{par:def}).

Applying (\ref{eq:rel}) to the second term of the action integral (\ref{eq:action2}) gives
$(\psi_{int}^0(t)| U[u](\xi_1,...,\xi_{N-1}) |\psi_{int}^0(t))=\frac{1}{2} \int \! d\vec{r} \, d\vec{r'} \gamma_{int}(\vec{r},\vec{r'};t) u(\vec{r}-\vec{r'})$,
where $\gamma_{int}$ is defined in (\ref{eq:gamint0}).

The action integral (\ref{eq:action2}) can thus be rewritten
\begin{eqnarray}
A_{int}[\rho_{int}] &=& \int_{t_0}^{t_1} dt (\psi_{int}^0(t)|i\hbar\partial_t-\sum_{\alpha=1}^{N-1} \frac{\tau_\alpha^2}{2\mu_\alpha}|\psi_{int}^0(t))
\nonumber\\
&& - \frac{1}{2} \int_{t_0}^{t_1} dt \int \! d\vec{r} \, d\vec{r'} \gamma_{int}(\vec{r},\vec{r'};t) u(\vec{r}-\vec{r'})
\nonumber\\
&& - \int_{t_0}^{t_1} dt \int \! d\vec{r} \; v_{int}(\vec{r};t) \rho_{int}(\vec{r},t)
.
\label{eq:action3}
\end{eqnarray}

Up to now we did not do any hypothesis.
To recover the \old{associated} Internal time-dependent KS scheme, we assume, as to obtain the
traditional time-dependent KS scheme \cite{Run84,Gro94}, that there exists, \textit{in the c.m.\ frame}, a $N$-body non-interacting system
(i.e. a local single-particle potential $v_S$)
%
\begin{eqnarray}
\label{eq:td_KS}
\Big( -\frac{\hbar^2\Delta}{2m} + v_S(\mathbf{r},t) \Big)\varphi^i_{int}(\mathbf{r},t) = i\hbar\partial_t \varphi^i_{int}(\mathbf{r},t)
\end{eqnarray}
which reproduces \textit{exactly} the density $\rho_{int}$ of the interacting system
(mind that $\rho_{int}$ is defined in the c.m. frame)
\begin{eqnarray}
\rho_{int}(\mathbf{r},t) =\sum_{i=1}^N |\varphi^i_{int}(\mathbf{r},t)|^2 .
\label{eq:rho_int}
\end{eqnarray}
%
Even if only ($N-1$) coordinates are sufficient to describe the internal properties,
they still describe a system of $N$ particles. Thus, we have to introduce $N$ orbitals in the KS scheme (as we did)
if we want them to be interpreted (to first order only) as single-particle orbitals
and obtain a scheme comparable (but not equivalent) to mean-field like calculations with effective interactions.

In (\ref{eq:td_KS}) we implicitely supposed that the particles are Fermions (a KS scheme to describe Boson condensates can be set similarly equalling all the $\varphi^i_{int}$).
Uniqueness of the potential $v_S(\mathbf{r},t)$ for a given density $\rho_{int}(\mathbf{r},t)$
(and initial $|\varphi^i_{int}(t_0))$ which yield the correct initial density $\rho_{int}(\mathbf{r},t_0)$)
is ensured by a direct application of the traditional time-dependent DFT formalism \cite{Run84,Gro94}.
Of course, the question of the validity of the KS hypothesis, known as the \textit{non-interacting v-representability} problem,
remains, as in traditional (time-dependent) DFT \cite{Dre90,Gro94}.

To use similar kinds of notations than the traditional DFT ones, we add and substract to the internal action integral (\ref{eq:action3}) the internal Hartree term
\\
$
A_{H}[\rho_{int}] = \frac{1}{2} \int_{t_0}^{t_1} dt \int \! d\vec{r} \, d\vec{r'} \, 
\rho_{int}(\vec{r},t) \, \rho_{int}(\vec{r'},t) \, u(\vec{r}-\vec{r'})
$,
the non-interacting kinetic energy term\\
$\int_{t_0}^{t_1} dt \sum_{i=1}^{N} (\varphi^i_{int}(t)|\frac{\vec{p}^2}{2m}|\varphi^i_{int}(t))$
and the
$\int_{t_0}^{t_1} dt \sum_{i=1}^{N} (\varphi^i_{int}(t)|i\hbar\partial_t|\varphi^i_{int}(t))$
term.
This permits to rewrite the ``internal'' action integral (\ref{eq:action3}) as
\begin{eqnarray}
A_{int}&=& \int_{t_0}^{t_1} dt \sum_{i=1}^{N} (\varphi^i_{int}(t)|i\hbar\partial_t-\frac{\vec{p}^2}{2m}|\varphi^i_{int}(t)) - A_{H}[\rho_{int}]
\nonumber\\
&& - A_{XC}[\rho_{int}] - \int_{t_0}^{t_1} dt \int \! d\vec{r} \; v_{int}(\vec{r};t) \, \rho_{int}(\vec{r},t) 
\label{eq:action4}
\end{eqnarray}
where the internal exchange-correlation part is defined as
\begin{widetext}
\begin{eqnarray}
A_{XC}[\rho_{int}]
&=& \frac{1}{2} \int_{t_0}^{t_1} dt \int \! d\vec{r} \, d\vec{r'} \, 
      \Big( \gamma_{int}(\vec{r},\vec{r'};t) - \rho_{int}(\vec{r},t) \, \rho_{int}(\vec{r'},t) \Big) \, 
      u(\vec{r}-\vec{r'})
\nonumber\\
&&    + \int_{t_0}^{t_1} dt \Big( 
      (\psi_{int}^0(t)|\sum_{\alpha=1}^{N-1} \frac{\tau_\alpha^2}{2\mu_\alpha}|\psi_{int}^0(t)) 
      - \sum_{i=1}^{N} (\varphi^i_{int}(t)|\frac{\vec{p}^2}{2m}|\varphi^i_{int}(t)) \Big)
\nonumber\\
&&    - \int_{t_0}^{t_1} dt \Big( (\psi_{int}^0(t)|i\hbar\partial_t|\psi_{int}^0(t))
      - \sum_{i=1}^{N} (\varphi^i_{int}(t)|i\hbar\partial_t|\varphi^i_{int}(t)) \Big)
.
\label{eq:Axc}
\end{eqnarray}
\end{widetext}
We see that it contains the exchange-correlation which comes from the interaction $u$ (first line of (\ref{eq:Axc})), but also the correlations contained in the interacting
kinetic energy  (second line of (\ref{eq:Axc})) and in the interacting ``$i\hbar\partial_t$'' term  (third line of (\ref{eq:Axc})).
A key point is that, as the KS assumption implies $\varphi^i_{int}[\rho_{int}]$ \cite{Run84,Gro94,Dre90},
$A_{XC}[\rho_{int}](t)$ can be written as a functional of $\rho_{int}$
(for given $|\psi_{int}^0(t_0))$ and $\{|\varphi^i_{int}(t_0))\}$ which yield the same initial density $\rho_{int}(\mathbf{r},t_0)$).

It remains to vary the ``internal'' quantum action (\ref{eq:action4}) to obtain the equations of motion (which define $\rho_{int}$).
Vignale, see Ref. \cite{Vig08}, showed recently that the correct formulation of the variational principle is not to stationarize the quantum action, i.e. $\delta A_{int}[\rho_{int}]=0$ as done so far \cite{ker76,Run84,Gro94}, but
\begin{eqnarray}
\delta A_{int}[\rho_{int}]=&&
i\big(\psi_{int}[\rho_{int}](t_1)\big|\delta \psi_{int}[\rho_{int}](t_1)\big)
\nonumber\\
&& - i \big(\psi^S_{int}[\rho_{int}](t_1)\big|\delta \psi^S_{int}[\rho_{int}](t_1)\big)
\label{eq:action}
\end{eqnarray}
(where $\psi^S_{int}$ is the Slater determinant constructed from the $\varphi^i_{int}$).
The two formulations lead to identical final results for theorems derived form symmetries of the action functional because
compensations occur \cite{Vig08}, but Vignales's formulation permits to solve the causality paradox of the previous formulation.

Varying (\ref{eq:action}) with respect to the $\varphi^{i*}_{int}(\mathbf{r},t)$, with $t\in[t_0,t_1]$,
leads straightforwardly to the Internal time-dependent KS equations for the $\varphi^i_{int}$
\begin{equation}
\label{eq:varphi_i}
\Big(
- \frac{\hbar^2}{2m}\Delta 
+ U_H[\rho_{int}] 
+ U_{XC}[\rho_{int}] 
+ v_{int}
\Big) \varphi^i_{int} = i\hbar\partial_t \varphi^i_{int}
\end{equation}
with the potentials
\begin{widetext}
\begin{eqnarray}
&& U_{H}[\rho_{int}](\vec{r},t) 
= \frac{\delta A_{H}[\rho_{int}]}{\delta \rho_{int}(\vec{r},t)}
\nonumber\\
&& U_{XC}[\rho_{int}](\vec{r},t) 
= \frac{\delta A_{XC}[\rho_{int}]}{\delta \rho_{int}(\vec{r},t)}
-i\big(\psi_{int}[\rho_{int}](t_1)\big|\frac{\delta \psi_{int}[\rho_{int}](t_1)}{\delta \rho_{int}(\vec{r},t)}\big)
+ i \big(\psi^S_{int}[\rho_{int}](t_1)\big|\frac{\delta \psi^S_{int}[\rho_{int}](t_1)}{\delta \rho_{int}(\vec{r},t)}\big)
\label{eq:Uxc}
\end{eqnarray}
\end{widetext}
which are local as expected ($v_S=U_H[\rho_{int}]+ U_{XC}[\rho_{int}] + v_{int}$ with the notations of Eq. (\ref{eq:td_KS})).
Note that the variational formulation of Vignale \cite{Vig08} leads to the addition of the last
two terms in the definition of $U_{XC}[\rho_{int}](\vec{r},t)$, see Eq. (\ref{eq:Uxc}),
compared to the traditional result obtained by stationarization of the action. It are those terms which permit to solve the causality paradox \cite{Vig08}.

Equations~(\ref{eq:varphi_i}) have the same form as the traditional time-dependent KS 
equations formulated \old{in the laboratory frame} for non-translationally 
invariant Hamiltonians \cite{Koh65,Run84,Gro94}
and permit to define $\rho_{int}$ through (\ref{eq:rho_int}).
Here, we have justified their form \textit{in the c.m. frame} for 
self-bound systems described with translationally invariant Hamiltonians.

But there is a major difference with the traditional DFT formalism.
Following similar steps as in Eq.~(\ref{eq:rel}), one can show that the
interacting kinetic energy term and the interacting ``$i\hbar\partial_t$'' term can be rewritten \cite{foot3}
\begin{widetext}
\begin{eqnarray}
&& (\psi_{int}^0(t)|\sum_{\alpha=1}^{N-1} \frac{\tau_\alpha^2}{2\mu_\alpha}|\psi_{int}^0(t))
= \int d\vec{r}_1 \cdots d\vec{r}_N \delta(\mathbf{R}) 
\psi_{int}^{0*}(\vec{r}_1,...,\vec{r}_N;t)
\sum_{i=1}^N \frac{\mathbf{p}_i^2}{2m}\psi_{int}^0(\vec{r}_1,...,\vec{r}_N;t)
\nonumber\\
&& (\psi_{int}^0(t)|i\hbar\partial_t|\psi_{int}^0(t)) =
\int d\vec{r}_1 \cdots d\vec{r}_N \delta(\mathbf{R}) \psi_{int}^{0*}(\vec{r}_1,...,\vec{r}_N;t)
i\hbar\partial_t \psi_{int}^0(\vec{r}_1,...,\vec{r}_N;t)
,
\label{eq:cm_cor}
\end{eqnarray}
\end{widetext}
which makes it clear that the differences with the non-interacting kinetic energy term
$\sum_{i=1}^{N} \int d\vec{r} \varphi^{i*}_{int}(\vec{r})\frac{\vec{p}^2}{2m}\varphi^i_{int}(\vec{r})$
and the non-interacting ``$i\hbar\partial_t$ term'' $\sum_{i=1}^{N} (\varphi^i_{int}(t)|i\hbar\partial_t|\varphi^i_{int}(t))$
(found in the exchange-correlation functional (\ref{eq:Axc}))
come, on the one hand, from the correlations neglected in the traditional independent-particle framework,
but also from the c.m. correlations described by the $\delta(\mathbf{R})$ term in (\ref{eq:cm_cor}), which does not appear in traditional time-dependent DFT \cite{Run84,Gro94}.
The inclusion of the c.m. correlations in the exchange-correlation functional (\ref{eq:Axc}) and potential (\ref{eq:Uxc})
is the main difference with the traditional KS scheme, and is a key issue for self bound-systems as atomic nuclei.

Mind that all the previous considerations only hold for fixed initial states $|\psi_{int}(t_0))$ and $\{|\varphi^i_{int}(t_0))\}$ which should of course give the same initial density $\rho_{int}(\mathbf{r},t_0)$ (and also for a fixed type of particle).
As a consequence, $\psi_{int}^0$ is not only a functional of $\rho_{int}$, but also depends on the initial state $|\psi_{int}(t_0))$, and $U_{XC}$, Eq. (\ref{eq:Uxc}),
also depends on the initial orbitals $\{|\varphi^i_{int}(t_0))\}$.
An important difference to the ground state Internal DFT formalism / KS scheme presented in \cite{Mes09}
is that $|\psi_{int}(t_0))$ and the $\{|\varphi^i_{int}(t_0))\}$
cannot necessarily be written as functionals of $\rho_{int}(\mathbf{r},t_0)$.
However, as underlined in \cite{Run84,Gro94}, if one starts from initial states $|\psi_{int}(t_0))$ and $\{|\varphi^i_{int}(t_0))\}$
that are non-degenerate ground states, i.e. that can be written as functionals of $\rho_{int}(\mathbf{r},t_0)$ \cite{Mes09},
$\psi_{int}$ and $U_{XC}$ become functionals of $\rho_{int}(\mathbf{r},t)$ alone.
Then, in the limit of stationary ground states, the theory reduces to the stationary Internal DFT / KS.

We recall that, as in traditional DFT, the previously discussed functionals are defined only for internal densities $\rho_{int}$ which correspond to some internal potential $v_{int}$, called \textit{v-representable} internal densities \cite{Run84,Gro94}.
Up to now, we do not know exactly how large the set of v-representable densities is.
This has to be kept in mind when variations with arbitrary densities are done, as to obtain the time-dependent KS equations.

\section{Conclusion.}

In summary, we have shown that, for a fixed initial state, the internal wave function,
which describes the internal properties of a time-dependent self-bound system, can be written (up to a trivial phase) as a functional of the internal density.
This implies that the "internal" expectation values of any observable (which does not contain a time derivative),
that are of experimental interest, can be regarded as functionals of the internal density.
Then, we set up, in the c.m.\ frame, a practical scheme which permits to calculate the internal density and whose form is similar to the traditional time-dependent KS equations,
the difference being that the exchange-correlation functional contains the c.m. correlations.

This work is a first step towards the justification to the use of density functionals
for time-dependent nuclear mean-field like calculations with effective interactions \cite{tdhf_nucl,Neg82},
proving that there exists an ultimate functional which permits to reproduce the exact internal density
(up to the non-interacting v-representability question).
If this functional was known, there would be no need for a c.m.\ correction.

Practically speaking, the time-dependent Internal KS scheme can describe, for instance in the nuclear case,
the collision of two nuclei in the frame attached to the total c.m.\ of the nuclei.
Then, $v_{int}$ is zero but the dependency to the initial state allows to start from
a state which corresponds to two nuclei with different velocities, or ``boosts''
(choosen such as the total kinetic momentum is zero because we are in the c.m. frame). According to the choice of the boosts, 
we can describe a wide variety of physical phenomena, from nuclear fusion \cite{tdhf_nucl} to
Coulomb excitation \cite{Ald56}. One of the nuclei can also simply consist in a particule as a proton,
to describe the excitation of a nucleus by diffusion.

A case where a non-zero $v_{int}$ would be interesting could be
the case of the laser irradiation ($v_{int}$ would then contain a laser potential switched on at $t>t_0$).
This is not of major interest in the nuclear case because, experimentally speaking,
we do not yet have lasers that are suited to the study of the laser irradiation of a nucleus.
However, this could be interesting in view of a generalization of this work to the whole molecule
(following from the generalization to different types of particules, which is underway).

Many questions remain open.
In particular, the question of the form of the potential which describes the c.m.\ correlations;
in addition to its practical interest, this question would also give interesting arguments concerning the non-interacting v-representability question.
Generalization to different types of particles (Fermions or Bosons) appears desirable.
Finally, the same reasoning should be applied
to rotational invariance to formulate the theory in term of the so-called "intrinsic" one-body density \cite{Gir08a}
(which is not directly observable). This is more complicated because rotation
does not decouple from internal motion, but it should be interesting concerning the symmetry breaking question.

\begin{acknowledgments}

The author is particularly grateful to M. Bender, E.K.U. Gross, and E. Suraud for enlightening discussions and
reading of the manuscript,
and thanks the referee for his pertinent remarks.
The author thanks the Centre d'Etudes Nucl\'eaires de Bordeaux-Gradignan for warm hospitality,
and the Institut Universitaire de France and the
Agence Nationale de la Recherche (ANR-06-BLAN-0319-02) for financial support.  


\end{acknowledgments}


\appendix

\section{Expression of the $\{\mathbf{r}_i-\mathbf{R}\}$ as functions of the $\{\xi_\alpha\}$ coordinates.}
\label{app:jacobi}

Using the relations (\ref{eq:jacobi}), one could show that the $\{\mathbf{r}_i-\mathbf{R}\}$ can all be
written in function of the $\{\xi_\alpha\}$ coordinates. We obtain, as a result
\begin{eqnarray}
&&\mathbf{r}_N-\mathbf{R}=\frac{N-1}{N}\xi_{N-1} ,
\nonumber\\
&&\mathbf{r}_{N-1}-\mathbf{R}=\frac{N-2}{N-1}\xi_{N-2} - \frac{1}{N}\xi_{N-1} ,
\nonumber\\
&&\mathbf{r}_{N-2}-\mathbf{R}=\frac{N-3}{N-2}\xi_{N-3} - \frac{1}{N-1}\xi_{N-2} - \frac{1}{N}\xi_{N-1},
\nonumber\\
&& \quad\vdots
\nonumber\\
&& \forall i \in [2,N-2]:
\nonumber\\
&& \quad\quad \mathbf{r}_{i}-\mathbf{R}=\frac{i-1}{i}\xi_{i-1} - \sum_{\alpha=i}^{N-2} \frac{1}{\alpha+1}\xi_{\alpha} - \frac{1}{N}\xi_{N-1},
\nonumber\\
&& \quad\vdots
\nonumber\\
&& \mathbf{r}_{1}-\mathbf{R}= - \sum_{\alpha=1}^{N-2} \frac{1}{\alpha+1}\xi_{\alpha} - \frac{1}{N}\xi_{N-1}
.
\nonumber
\end{eqnarray}
The $\xi_{N-1}$ is the only Jacobi coordinate to appear in the expressions (in function of the Jacobi coordinates) of \textit{all} the $\{\mathbf{r}_i-\mathbf{R}\}$.
Thus, all the $v_{int}(\mathbf{r}_i-\mathbf{R};t)$ terms that enter into the calculation of the total internal potential, Eq. (\ref{eq:V_int}), 
will contain $\xi_{N-1}$.

\section{Expression of the internal probability current $\mathbf{j}_{int}(\mathbf{r},t)$.}
\label{app:int_current}

The one-body total laboratory density and probability current are defined as
\begin{eqnarray}
\rho(\mathbf{r},t)&=& N \int d\mathbf{r}_1...d\mathbf{r}_{N-1}|\psi(\mathbf{r}_1,...,\mathbf{r}_{N-1},\mathbf{r};t)|^2 \nonumber\\
\nonumber\\
\mathbf{j}(\mathbf{r},t)&=&\frac{\hbar}{2m i} N \int d\mathbf{r}_1...d\mathbf{r}_{N-1} \psi^*(\mathbf{r}_1,...,\mathbf{r}_{N-1},\mathbf{r};t)
\nonumber\\
&&\hspace{1.3cm}\times \mathbf{\nabla_r}\psi(\mathbf{r}_1,...,\mathbf{r}_{N-1},\mathbf{r};t)  + c.c.
\end{eqnarray}
They satisfy the ``laboratory'' continuity equation
\begin{eqnarray}
\partial_t \rho(\mathbf{r},t) + \mathbf{\nabla_r} . \mathbf{j}(\mathbf{r},t) =0 .
\label{eq:cr}
\end{eqnarray}
We do some manupulations on the laboratory probability current $\mathbf{j}$ using the Jacobi coordinates ($c.c.$ denotes the complex conjugate; the second equality is obtained using $\mathbf{\nabla_{r_N}} = \mathbf{\nabla_{\xi_{N-1}}} + \mathbf{\nabla_{R}}/N$, by definition of the Jacobi coordinates)
\begin{widetext}
\begin{eqnarray}
\mathbf{j}(\mathbf{r},t)&=&\frac{\hbar}{2m i} N \int d\mathbf{r}_1...d\mathbf{r}_{N-1} d\mathbf{r}_N \delta (\mathbf{r}-\mathbf{r}_N) \psi^*(\mathbf{r}_1,...,\mathbf{r}_{N-1},\mathbf{r}_N;t) \mathbf{\nabla_{r_N}}\psi(\mathbf{r}_1,...,\mathbf{r}_{N-1},\mathbf{r}_N;t)  + c.c.  
\nonumber\\
&=& \frac{\hbar}{2m i} N \int d\mathbf{R} d\mathbf{\xi}_1 ... d\mathbf{\xi}_{N-1} \Big( \frac{N}{N-1} \Big)^3 \delta\Big(\xi_{N-1}-\frac{N}{N-1}(\mathbf{r}-\mathbf{R})\Big)  
\nonumber\\
&& \times \Gamma^*(\mathbf{R},t) \psi_{int}^*(\mathbf{\xi}_1, ..., \mathbf{\xi}_{N-1};t) \Big( \mathbf{\nabla_{\xi_{N-1}}} + \frac{\mathbf{\nabla_{R}}}{N} \Big) \Gamma(\mathbf{R},t) \psi_{int}(\mathbf{\xi}_1, ..., \mathbf{\xi}_{N-1};t) + c.c.  
\nonumber\\
&=& \int d\mathbf{R} |\Gamma(\mathbf{R},t)|^2  
\times \frac{\hbar}{2m i} N  \Big( \frac{N}{N-1} \Big)^3 \int d\mathbf{\xi}_1 ...  d\mathbf{\xi}_{N-2} \psi_{int}^*(\mathbf{\xi}_1, ..., \mathbf{\xi}_{N-2}, \nu;t) \mathbf{\nabla_{\nu}} \psi_{int}(\mathbf{\xi}_1, ..., \mathbf{\xi}_{N-2}, \nu;t) \Big|_{\nu=\frac{N}{N-1}(\mathbf{r}-\mathbf{R})} 
\nonumber\\
&& + \frac{\hbar}{2m i} \int d\mathbf{R} \Gamma^*(\mathbf{R},t) \frac{\mathbf{\nabla_{R}}}{N} \Gamma(\mathbf{R},t) 
\times N \Big( \frac{N}{N-1} \Big)^3 \int d\mathbf{\xi}_1 ...  d\mathbf{\xi}_{N-2} |\psi_{int} (\mathbf{\xi}_1, ..., \mathbf{\xi}_{N-2},\frac{N}{N-1}(\mathbf{r}-\mathbf{R});t)|^2 + c.c.  
\nonumber\\
&=& \int d\mathbf{R} |\Gamma(\mathbf{R},t)|^2 \mathbf{j}_{int}(\mathbf{r}-\mathbf{R},t) + \int d\mathbf{R} \rho_{int}(\mathbf{r}-\mathbf{R},t) \mathbf{j}_{\Gamma}(\mathbf{R},t) ,
\label{eq:j_int2}
\end{eqnarray}
where we introduced the internal one-body density (\ref{eq:rho_int0}), the c.m.\ probability current
$\mathbf{j}_{\Gamma}(\mathbf{R},t) = \frac{\hbar}{2 M i} \Gamma^*(\mathbf{R},t) \mathbf{\nabla_{R}} \Gamma(\mathbf{R},t) + c.c.$
($M=Nm$ is the total mass), and the internal probability current (\ref{eq:j_int1}).
The meaning of Eq. (\ref{eq:j_int2}) is clear: the laboratory probability current is the sum of the c.m.\ probability current and of the internal probability current,
both convolued respectively with the internal one-body density and the c.m.\ one-body density.
One can show that $\mathbf{j}_{\Gamma}$ and $\mathbf{j}_{int}$ satisfy both independent continuity relations.
It is trivial, using (\ref{eq:H_cm}) and (\ref{eq:schro_int}), for $\mathbf{j}_{\Gamma}$. For $\mathbf{j}_{int}$, we calculate (with the help of (\ref{eq:H_int}) and (\ref{eq:schro}))
\begin{eqnarray}
&& \Big( \frac{N}{N-1} \Big)^3 \partial_t \int d\mathbf{\xi}_1 ...  d\mathbf{\xi}_{N-1} \delta(\mathbf{\xi}_{N-1}-\frac{N}{N-1}\mathbf{r}) |\psi_{int}(\mathbf{\xi}_1, ..., \mathbf{\xi}_{N-1} ;t)|^2
\nonumber\\
&& \hspace{1cm} = - \frac{1}{i\hbar} \Big( \frac{N}{N-1} \Big)^3\int d\mathbf{\xi}_1 ...  d\mathbf{\xi}_{N-1} \delta(\mathbf{\xi}_{N-1}-\frac{N}{N-1}\mathbf{r})
\psi_{int}^*(\mathbf{\xi}_1, ..., \mathbf{\xi}_{N-1} ;t)  \frac{\hbar^2}{2\mu_{N-1}}\Delta_{\xi_{N-1}} \psi_{int}(\mathbf{\xi}_1, ..., \mathbf{\xi}_{N-1} ;t) + c.c.
\nonumber\\
&& \hspace{1cm} = - \frac{\hbar}{2 \mu_{N-1} i} \Big( \frac{N}{N-1} \Big)^3 \int d\mathbf{\xi}_1 ...  d\mathbf{\xi}_{N-2} \psi_{int}^*(\mathbf{\xi}_1, ...,\mathbf{\xi}_{N-2}, \nu ;t) \Delta_{\nu} \psi_{int}(\mathbf{\xi}_1, ...,\mathbf{\xi}_{N-2}, \nu ;t)\Big|_{\nu=\frac{N}{N-1}\mathbf{r}} + c.c.
\nonumber\\
&& \hspace{1cm} = - \frac{\hbar}{2 \mu_{N-1} i} \Big( \frac{N}{N-1} \Big)^3 \nabla_{\nu} . \int d\mathbf{\xi}_1 ...  d\mathbf{\xi}_{N-2} \psi_{int}^*(\mathbf{\xi}_1, ...,\mathbf{\xi}_{N-2}, \nu ;t) \nabla_{\nu} \psi_{int}(\mathbf{\xi}_1, ...,\mathbf{\xi}_{N-2}, \nu ;t)\Big|_{\nu=\frac{N}{N-1}\mathbf{r}} + c.c.
\nonumber\\
&& \hspace{1cm} = - \frac{\hbar}{2 m i} \Big( \frac{N}{N-1} \Big)^3 \nabla_{\mathbf{r}} . \int d\mathbf{\xi}_1 ...  d\mathbf{\xi}_{N-2} \psi_{int}^*(\mathbf{\xi}_1, ...,\mathbf{\xi}_{N-2}, \nu ;t) \nabla_{\nu} \psi_{int}(\mathbf{\xi}_1, ...,\mathbf{\xi}_{N-2}, \nu ;t)\Big|_{\nu=\frac{N}{N-1}\mathbf{r}} + c.c.
\nonumber
\end{eqnarray}
\end{widetext}
(To obtain the last equality we used the fact that, by definition, $\mu_{N-1}=\frac{N-1}{N}m$.)
From this relation we deduce, using (\ref{eq:rho_int0}) and (\ref{eq:j_int1}), the ``internal'' continuity equation
\begin{eqnarray}
\partial_t \rho_{int}(\mathbf{r},t) + \mathbf{\nabla_\mathbf{r}} . \mathbf{j}_{int}(\mathbf{r},t) =0 .
\nonumber
\end{eqnarray}
This reinforces the interpretation of $\mathbf{j}_{int}$ as the internal probability current.

\section{Integral mean value theorem for functions of many variables.}
\label{app:mean_val_th}

We give the generalization of the
mean value theorem \cite{bass} to functions of an arbitrary number of variables.
One starts from two real functions of A variables
\begin{eqnarray}
\forall (x_1,...,x_A) \in \Re e : \hspace{5mm} && f : [x_1,...,x_A]\mapsto \Re e
\nonumber\\
&& g : [x_1,...,x_A]\mapsto \Re e .
\nonumber
\end{eqnarray}
We suppose that they are integrable in a domain $D$, that $f\ge 0$ in $D$, and that $g$ is continuous in $D$. We define
\begin{eqnarray}
m=inf\{g(x_1,...,x_A) ; (x_1,...,x_A)\in D\}
\nonumber\\
M=sup\{g(x_1,...,x_A) ; (x_1,...,x_A)\in D\}.
\nonumber
\end{eqnarray}
As $f\ge 0$, $m f \le f g \le M f$, which we integrate
\begin{widetext}
\begin{eqnarray}
&& m \int_D dx_1...dx_A f(x_1,...,x_A) \le \int_D dx_1...dx_A f(x_1,...,x_A) g(x_1,...,x_A) \le M \int_D dx_1...dx_A f(x_1,...,x_A)
\nonumber\\
&& \Rightarrow \exists C \in [m,M] : \int_D dx_1...dx_A f(x_1,...,x_A) g(x_1,...,x_A) = C \int_D dx_1...dx_A f(x_1,...,x_A) .
\nonumber
\end{eqnarray}
As we supposed that $g$ is continuous in $D$, we deduce that
$\exists (x'_1,...,x'_A) \in D$ : $g(x'_1,...,x'_A)=C$,
which implies
\begin{eqnarray}
\exists (x'_1,...,x'_A) \in D : \int_D dx_1...dx_A f(x_1,...,x_A) g(x_1,...x_A) = g(x'_1,...,x'_A) \int_D dx_1...dx_A f(x_1,...,x_A) .
\nonumber
\end{eqnarray}
\end{widetext}

\bibliography{apssamp}

\begin{thebibliography}{99}
\bibitem{foot1}
Translational invariance of the wave function,
which implies that the mean values of the observables do not depend on the position of the c.m., 
is a weaker condition than the more fundamental
Galilean invariance of the wave function \cite{Schm01a}.
\bibitem{foot2}
A multicomponent DFT formalism, for an explicit quantum treatment of both the electrons and the nuclei of a molecule,
has been proposed in Ref. \cite{Kre01,Kre08}.
\bibitem{foot3}
More generally, we should introduce a $\delta(\mathbf{R}-\mathbf{a})$, where
$\mathbf{a}$ is an arbitrary translation vector, instead of the $\delta(\mathbf{R})$.
This is linked to the translational invariance.
In this paper we chose $\mathbf{a}=\mathbf{0}$, which leads to perfectly equivalent results
and permits to lighten the notations.
\end{thebibliography}

\end{document}